%% file: mimo_ofdm_jrc_sdr_arxiv.tex
\documentclass[conference]{IEEEtran}
\IEEEoverridecommandlockouts

\usepackage{cite}
\usepackage{amsmath,amssymb,amsfonts}
\usepackage{algorithmic}
\usepackage{graphicx}
\usepackage{textcomp}
\usepackage{xcolor}
\def\BibTeX{{\rm B\kern-.05em{\sc i\kern-.025em b}\kern-.08em
		T\kern-.1667em\lower.7ex\hbox{E}\kern-.125emX}}

\usepackage{multirow}
\usepackage{upgreek}
\usepackage{tabularx}
\usepackage{threeparttable}
\usepackage[font=small]{caption}
\usepackage{subcaption}
\usepackage{amsthm}

\usepackage{enumitem}
\usepackage[ruled, linesnumbered]{algorithm2e}

\input{mycommands.tex}

\begin{document}

\title{Software-Defined MIMO OFDM Joint Radar-Communication Platform with Fully Digital mmWave Architecture \thanks{This work has been supported in part by NSF through grants 1814923, 1955535, 2030141, 2112471.}
}

\author{\IEEEauthorblockN{Ceyhun D. Ozkaptan, Haocheng Zhu, Eylem Ekici}
	\IEEEauthorblockA{Dept. of Electrical and Computer Engineering \\
		The Ohio State University\\
		Columbus, OH, USA \\
		\{ozkaptan.1, zhu.2991, ekici.2\}@osu.edu}
	\and
	\IEEEauthorblockN{Onur Altintas}
	\IEEEauthorblockA{InfoTech Labs,\\
		Toyota Motor North America \\
		Mountain View, CA, USA \\
		onur.altintas@toyota.com}
	}

\maketitle

\begin{abstract}

Large-scale deployment of connected vehicles with cooperative sensing and maneuvering technologies increases the demand for vehicle-to-everything communication (V2X) band in 5.9 GHz. Besides the V2X spectrum, the under-utilized millimeter-wave (mmWave) bands at 24 and 77 GHz can be leveraged to supplement V2X communication and support high data rates for emerging broadband applications. For this purpose, joint radar-communication (JRC) systems have been proposed in the literature to perform both functions using the same waveform and hardware. In this work, we present a software-defined multiple-input and multiple-output (MIMO) JRC with orthogonal frequency division multiplexing (OFDM) for the 24 GHz mmWave band. We implement a real-time operating full-duplex JRC platform using commercially available software-defined radios and custom-built mmWave front-ends. With fully digital MIMO architecture, we demonstrate simultaneous data transmission and high-resolution radar imaging capabilities of MIMO OFDM JRC in the mmWave band. 

\end{abstract}

\section{Introduction}
Vehicle-to-everything communication (V2X) and vehicular radar imaging have become the key enablers of Intelligent Transportation Systems (ITS) to improve coordination, safety, and automation in traffic. For V2X, Federal Communications Commission (FCC) dedicated 30 MHz of bandwidth in the 5.9 GHz band for the ITS applications and the exchange of safety-related messages. With the large-scale deployment of connected vehicles, the V2X-dedicated band will face a spectrum scarcity problem. For instance, emerging full self-driving, cooperative sensing, and maneuvering technologies require a large amount of sensor and navigation data to be exchanged for improved reliability and performance \cite{2015_cooperative_perception}. However, due to limited bandwidth, the V2X spectrum cannot support these broadband applications along with basic safety messages that are crucial for safety-critical applications \cite{5gaa_spectrum_needs_2020}.

A solution to alleviate the scarcity problem and attain higher data rates is to leverage the 250 MHz ISM band in the 24-24.25 GHz mmWave spectrum. While the 24 GHz ISM band allows unlicensed access, it can be utilized more effectively with vehicular joint radar-communication (JRC) systems that offer simultaneous radar imaging and data transmission using the same RF signal and transceiver. In particular, a JRC system promotes the efficient utilization of the spectrum while reducing power consumption, cost, and hardware size. 

In the literature, various experimental JRC testbeds have been proposed in \cite{sturm_ofdm_2011, braun2012usrp, ozkaptan_sdr_demo_2019} with single-input single-output (SISO) architectures with range-velocity estimation. Due to hardware and cost limitations, MIMO JRC systems are often studied via numerical simulations \cite{fan_liu_joint_design_2020}. Besides, an OFDM-based MIMO radar testbed has been presented in \cite{7811637} as an offline measurement setup without communication capability by assigning different subcarriers to different transmit antennas. In addition, a single-input multiple-output (SIMO) JRC setup has been evaluated in \cite{kumari_simo_jcr80_2021} with static scenarios. Nevertheless, the SIMO radar processing capability is achieved synthetically by sliding a transmit antenna. Hence, the sliding mechanism generates \textit{non-coherent} radar images that are digitally combined, which limits its feasibility for mobile scenarios.  

In our previous work \cite{9589830}, we proposed a MIMO-OFDM-based JRC framework that leverages all of the spatial and spectral resources for simultaneous MIMO radar imaging and multi-user communication. Moreover, we show that the proposed system generates range-angle images using reflected preamble and precoded waveform via numerical simulations. In this work, we present a proof-of-concept implementation of the MIMO OFDM JRC to demonstrate and evaluate its radar and communication performance. To the best of our knowledge, this is the first software-defined mmWave MIMO JRC platform that can perform real-time MIMO radar imaging and precoded data transmission using MIMO OFDM waveform. To make the platform accessible for other researchers, we release the implementation as open-source.\footnote{Available at https://github.com/ceyhunozkaptan/gr-mimo-ofdm-jrc.} The main contributions of this work are summarized as follows:
\begin{itemize}[leftmargin=*]
	\item  We present a software-defined mmWave MIMO OFDM transceiver for demonstrating and evaluating JRC capabilities with a fully-digital architecture with 4 transmit and 2 receive chains. To operate in the 24 GHz band, we also build custom mmWave front-ends using commercially available RF components for up/down-conversion
	\item We propose a self-interference cancellation method to remove direct leakage and reflections from static background for improved radar imaging performance.
	\item To demonstrate the range-angle imaging capabilities, we conduct mobile experiments using octahedral reflectors for single and two-target scenarios. Simultaneously, we evaluate the communication performance of the JRC platform at different distances using a separate SISO transceiver. 
\end{itemize}

\section{Hardware Architecture}

In this section, we present the hardware architecture that enables the implementation of digital signal processing (DSP) stages of the software-defined JRC testbed. For better flexibility in development, we employ USRP N32x series software-defined radios (SDRs) from National Instruments (NI) that provide host PC-based baseband processing capability along with FPGA-controlled sub-6 GHz front-ends. As illustrated in Figure~\ref{fig_hardware_schematic}, the MIMO JRC transceiver uses USRP N321 and N320 to obtain up to 4 transmit (TX) and 4 receive (RX) channels. Both USRP N320 and N321 provide the same capabilities as 200 MHz bandwidth and real-time IQ streaming with 10 Gbps Ethernet. The difference is that N321 can export its local oscillator (LO) signal, whereas N320 can import an external LO signal. The direct LO sharing between two USRPs allows RF chains to have \textit{deterministic} phase differences. Hence, after digital phase compensations, all chains perform phase-coherent analog-to-digital (ADC) conversions. 

 Aside from TX waveform samples, the SDRs with multiple RX channels generate around 3.2 GB/s of raw samples to be transferred and processed by the CPU of the Host PC. To sustain the high-rate streaming, the Host PC is equipped with three 10 Gbps Ethernet cards: two Intel X520-DA2 and one Intel X710-DA2 that provide six optical SFP+ ports in total. Moreover, the Host PC is required to complete DSP tasks with high throughput and low processing delay to enable real-time operations without buffer overflows or underruns. Therefore, the Host PC is equipped with 48 GB of memory and Intel i9-12900K 16-core CPU  with up to 5.2 GHz frequency.

\subsection{mmWave Front-end Design}

As an extension to USRPs, we designed a custom mmWave front-end unit that provides up and down conversion stages between intermediate frequency (IF) (i.e., sub-6 GHz) and RF (i.e., 24 GHz) band. In each TX chain, a transmit signal generated from the USRP goes through an upconverter, a band-pass filter (BPF), and a power amplifier. Oppositely in each RX chain, a received signal passes through a BPF, low-noise amplifier, a downconverter, and a low-pass filter. We employ the BPFs for image rejection on the lower sideband after or before the mixers based on the direction of conversion. With this architecture, we built 4 mmWave front-end units using modular RF components from X-Microwave.   

\begin{figure}[!t]
	\centering
	\includegraphics[width=\columnwidth]{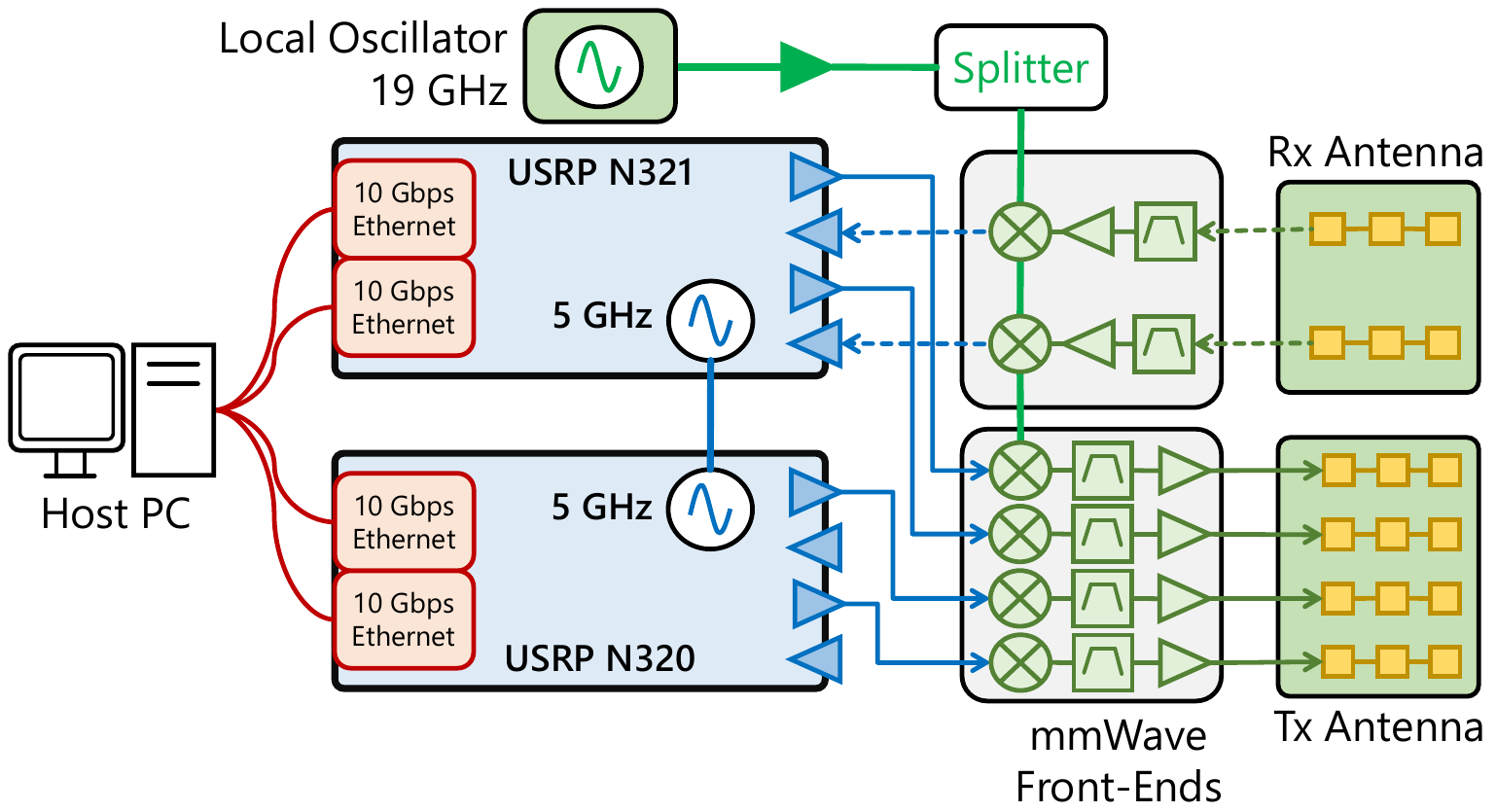} 
	\caption{Hardware architecture of MIMO OFDM JRC transceiver.}
	\label{fig_hardware_schematic}
\end{figure}

To maintain phase coherency in the conversion stage, we implement an LO distribution system using power splitters similar to the LO sharing mechanism of USRPs. As shown in Figure~\ref{fig_hardware_schematic}, a common LO input is shared between two identical chains via a power splitter. Moreover, the LO input is amplified before the power splitter to achieve the mixer's power requirement of 13 dBm. To compensate for the losses in mixing and image rejection stages, we also use power and low-noise amplifiers in transmit and receive chains, respectively. Table~\ref{table_mmwave_frontend} summarized the details of the RF components used in the mmWave front-ends. 

\begin{figure}[!b]
	\centering
	\includegraphics[width=\columnwidth]{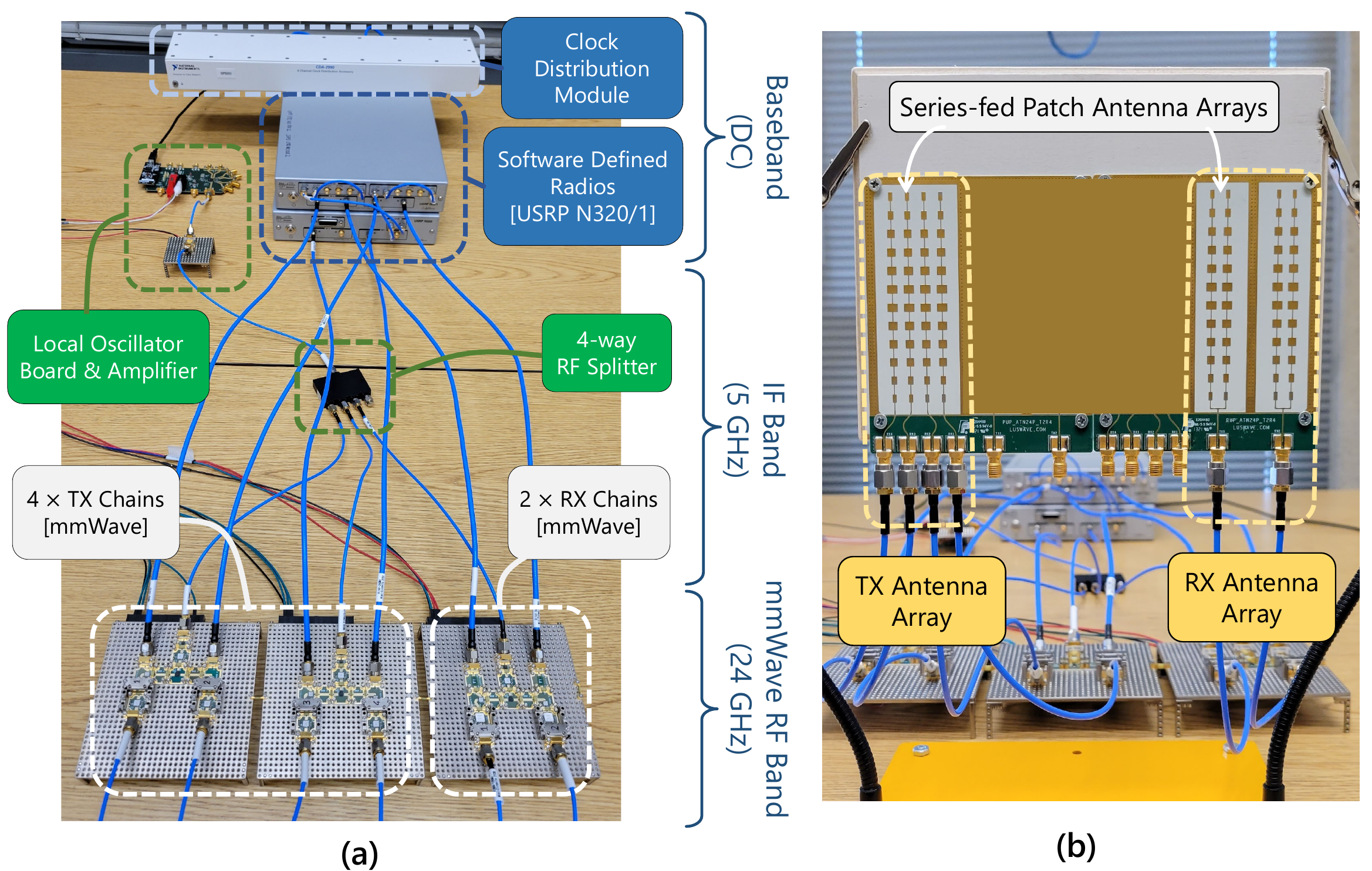} 
	\caption{Hardware architecture of 4$ \times $2 MIMO JRC transceiver.}
	\label{fig_bs_hardware}
\end{figure}

\subsection{MIMO JRC Transceiver Setup}

Since the MIMO JRC platform uses two separate USRPs, we employ a clock distribution module to share a common 10 MHz reference and pulse-per-second (PPS) signal to synchronize their internal clocks and timers. Hence, the control commands from the Host PC can be handled synchronously using timestamps.  As shown in Figure~\ref{fig_bs_hardware}-(a), a common LO signal at 19 GHz is generated from a wideband frequency synthesizer EVAL-ADF4371. To distribute the common LO signal to each mmWave front-end unit, we use a driver amplifier and a 4-way splitter. The MIMO JRC transceiver has 4 TX and 2 RX chains in total with a common LO.

We use a MIMO antenna array from Luswave Technology with $ \Ntx= $ 4 TX and $ \Nrx= $ 2 RX channels which is shown in Figure~\ref{fig_bs_hardware}-(b). Each TX channel is connected to \textit{a series-fed patch antenna array}, whereas each RX channel is connected to two parallel arrays for increased gain. Based on the specifications, each series-fed antenna array generates a fixed narrow beampattern along the vertical axis with 8 dBi gain. Moreover, each TX array is separated by $ d_{\text{tx}} = 6.35 $ mm, whereas each RX array is separated by  $ d_{\text{rx}} = 4d_{\text{tx}} = 25.4$ mm. 

\begin{table}[!t]
	\centering
	\resizebox{0.85\columnwidth}{!}{
		\begin{threeparttable}
			\begin{tabularx}{\columnwidth}{c c Y c }
				\hline \hline
				\multicolumn{4}{c}{\textbf{mmWave Front-end Components}}  \\ \hline
				\textbf{Component} & \textbf{Part Number} & \textbf{Frequency Range} & \textbf{Gain}\tnote{(1)}    \\ \hline
				Mixer & CMD179C3 & 16 - 26 GHz & -6.5 dB  \\
				Power and LO Amplifier & HMC498LC4 & 17 - 24 GHz & 21.9 dB \\
				Low-Noise Amplifier & HMC7950 & 2 - 28 GHz & 16.0 dB  \\
				Band Pass Filter & MFB-2625  & 21 - 30 GHz & -1.4 dB  \\ 
				Low Pass Filter & LP0AA6160A7 & DC - 6 GHz & -0.9 dB  \\
				2-way Splitter & EP2KA+ & 10 - 43 GHz & -4.2 dB  \\
				\hline\hline
			\end{tabularx}
			\begin{tablenotes}
				\item[(1)] Expected gains based on specifications and operating frequency.
			\end{tablenotes}
		\end{threeparttable}
	}
	\caption{Details of RF components used in the mmWave front-ends.}
	\label{table_mmwave_frontend}
\end{table}

\section{Software Architecture}\label{section_software_dsp} 

In this section, we present the software architecture of the MIMO OFDM-based JRC testbed that handles of four baseband operations: (i) MIMO front-end calibration, (ii) transmit waveform generation, (iii) MIMO OFDM radar processing, and (iv) communication receiver. While USRPs perform ADC/DAC with embedded front-ends and FPGAs, the baseband DSP is executed by the CPU of the Host PC. For real-time DSP, we use the GNU Radio framework, which is an open-source development platform for software-defined DSP blocks and radios. Using the framework, we implemented our DSP blocks in C++ along with already available blocks such as FFT, UDP Socket. For parallel processing, GNU Radio framework governs its own scheduler to utilize all CPU threads and internal buffers dynamically. Nevertheless, inefficient DSP blocks often cause bottlenecks and latency that lead to buffer overflows. Hence, one of our goals was to achieve high efficiency and avoid computationally expensive operations. In particular, we extensively resort to \textsf{Eigen 3}, an efficient C++ library for linear algebra operations \cite{eigenweb}. 

\subsection{MIMO OFDM Waveform Generation}

\begin{figure}[!t]
	\centering
	\includegraphics[width=\columnwidth]{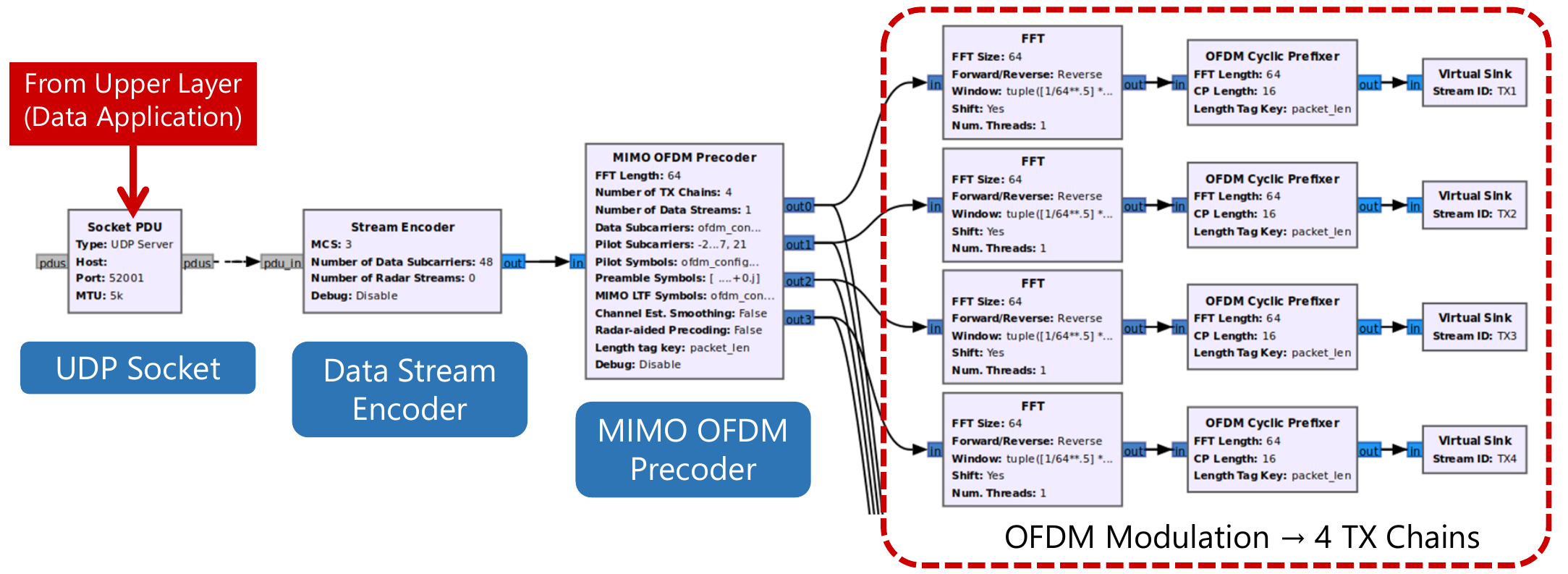} 
	\caption{Flowgraph of joint MIMO OFDM transmission with a single data stream received from a data application on upper layer.}
	\label{fig_mimo_ofdm_waveform}
\end{figure}

\textbf{Socket PDU:} Waveform generation starts with a Socket PDU block that provides a UDP socket to receive data packets generated from a \textit{data application} running along with the DSP flowgraph generated with GNU Radio. As illustrated in Figure~\ref{fig_mimo_ofdm_waveform}, once a data packet is received with the UDP block, it is forwarded to the Stream Encoder block. 

\textbf{Stream Encoder:} This block first computes a 32-bit cyclic redundancy check (CRC) of the received packet and appends it to the packet. Based on chosen modulation and coding scheme (MCS), the data bytes are encoded into bits after scrambling, convolution encoding, and puncturing operations. Finally, encoded bits are converted into complex values symbols with the chosen modulation scheme. The supported MCSs are BPSK, QPSK, QAM with coding rates of 1/2 and 3/4. Moreover, this block tags frames as a Null Data Packet (NDP) or Data Packet (DATA) to adjust the precoder accordingly.

\textbf{MIMO OFDM Precoder:} This block generates TX frames that are precoded to multiple transmit antennas. A TX frame consists of 2 short training symbols, 2 long training symbols, 1 header symbol, $ \Ntx $ MIMO preamble symbols, and precoded data symbols consecutively. The short and long training symbols are transmitted from the first 2 TX antennas without precoding to achieve a wider beampattern. These training symbols are used by a receiver for frame detection, time and frequency synchronization. For accurate MIMO channel estimation, NDPs only contain the orthogonal MIMO preamble without precoding. For DATA, both data and MIMO preamble symbols are precoded. In NDP-assisted precoding, the precoder block computes the steering matrix using the latest MIMO channel estimate that is feedbacked from the receiver. Finally, OFDM modulation is performed separately for each TX chain with IFFTs and cyclic prefixers. 

\begin{figure}[!b]
	\centering
	\includegraphics[width=\columnwidth]{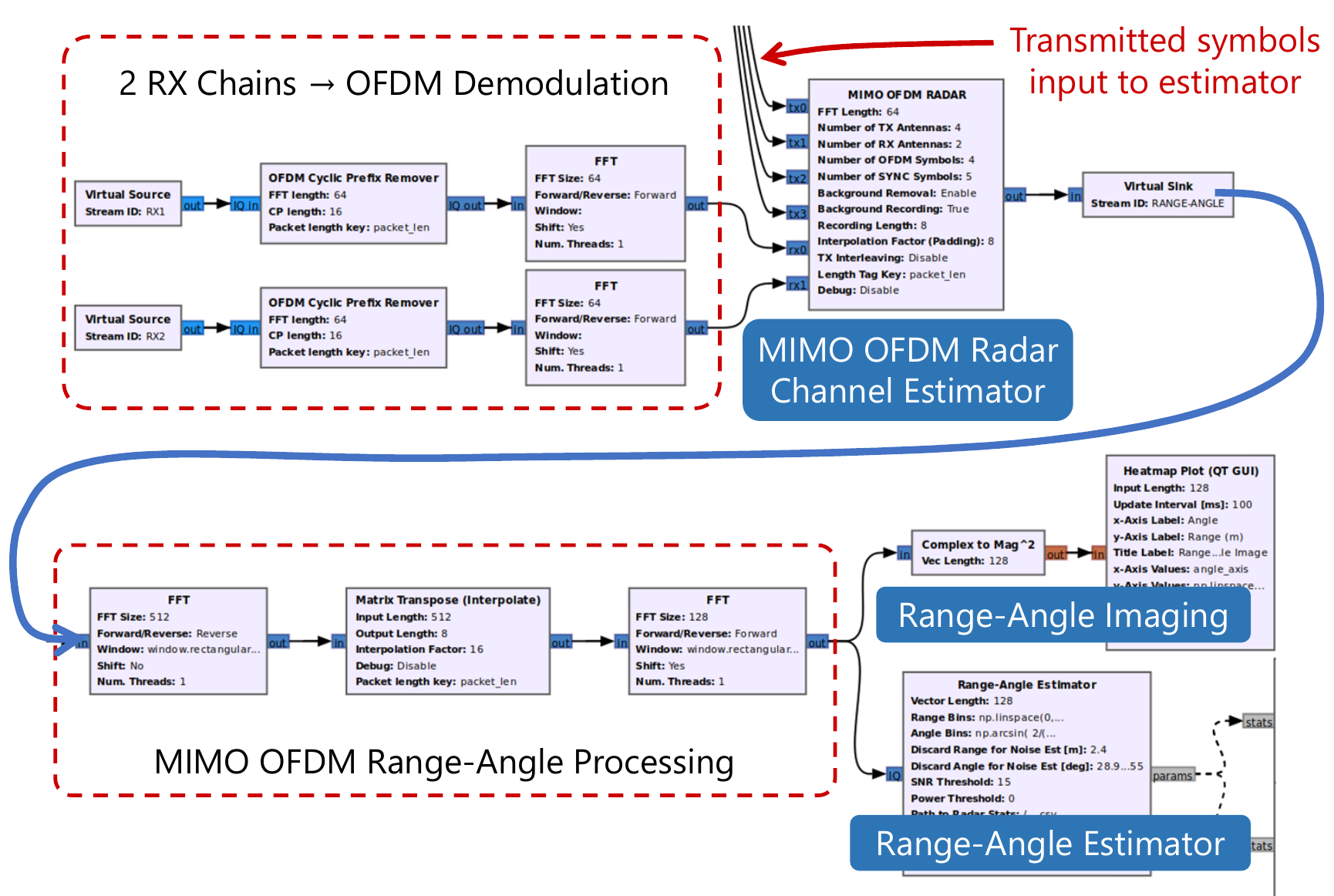} 
	\caption{Flowgraph of MIMO OFDM-based radar channel estimation and imaging operations after reception.}
	\label{fig_mimo_radar_image}
\end{figure}

\subsection{MIMO OFDM Radar Imaging}

During the transmission of the MIMO OFDM waveform, the JRC transceiver synchronously starts reception to obtain reflected symbols for radar imaging. Overall, Figure~\ref{fig_mimo_ofdm_waveform} and \ref{fig_mimo_radar_image} display the complete flowgraph of the baseband processing performed by the MIMO JRC transceiver from waveform generation to real-time radar imaging. 

\textbf{MIMO OFDM Radar:} This block follows the radar channel estimation algorithm proposed in \cite{9589830} after the OFDM demodulation is performed with FFT blocks as depicted in Figure~\ref{fig_mimo_radar_image}.  In addition, it keeps track of the static radar channel that contains self-interference and background reflection components to be removed from the radar image. The details of this approach are explained in Section~\ref{6_si_cancellation}. 

\textbf{Range-Angle Imaging:} After the unstructured channel matrix is obtained, the range-angle image generation algorithm in \cite{9589830} is executed with 2 FFT operations as shown in Figure~\ref{fig_mimo_radar_image}. Finally, range-angle images are displayed as heatmap figures in real-time, while extracted radar parameters (i.e., range, angle, SNR) are also tracked with time plots.

\subsection{Communication Receiver}

To demonstrate the communication capability with the same waveform, we also implement a communication receiver flowgraph that performs frame detection, synchronization, estimation, equalization, and demodulation stages. We develop our MIMO OFDM receiver based on a SISO OFDM (i.e., 802.11p) receiver implementation presented in \cite{bloessl_transceiver_2018}. 

\textbf{Frame Detection and Synchronization:}  To detect incoming frames, we employ the delay-and-correlate algorithm that is used in 802.11 Wi-Fi standards \cite{terry2002ofdm}. Due to false detection caused by DC offset, we also use a DC-blocking FIR filter before the correlator. Once a frame is detected, carrier frequency offset estimation is performed by leveraging the short training symbols in the time domain. 

\textbf{MIMO OFDM Equalizer:} Unlike other receiver blocks that are adopted from \cite{bloessl_transceiver_2018}, we implement our MIMO OFDM equalizer block that performs MIMO channel estimation, data stream equalization, and SNR estimation. When an NDP frame is received, this block performs MIMO channel estimation and stores estimated channel matrix in a file. 

When a precoded DATA frame is received, the MIMO channel estimate is updated in the equalization stage. For the channel estimation, we employ two different approaches: (i) Least Squares (LS) estimator, and (ii)  Spectral Temporal Averaging (STA) estimator\cite{5982455}, which uses a decision-directed approach to update the estimate. Moreover, in this block, we compute the SNR estimate of precoded symbols using known symbols pilot subcarriers. After this block, equalized data symbols are demodulated, de-scrambled, and decoded using the Viterbi decoder in the Stream Decoder block.

\section{Self-Interference and Background Removal}\label{6_si_cancellation}

For radar imaging, the MIMO JRC transceiver always operates in full-duplex mode to simultaneously receive reflected signals. With full-duplexity and closely-placed TX-RX chains, the received radar signal is dominated by self-interference (SI) consisting of (i) direct leakage from TX to RX chains, and (ii) reflections from nearby scatterers. Since the reflected target signal experiences higher attenuation, the SI and near-field reflections mask targets on the radar image and reduce the probability of detection due to lowered dynamic range.

In this work, we design a digital SI and background removal method due to its simplicity. A mean subtraction technique has been used with ground penetrating radars to highlight targets by suppressing ground reflections \cite{5550079}. In this work, we implement an extended version of the mean subtraction method for MIMO OFDM radar in which it is applied to each virtual spatial channel separately in the frequency domain. 

After the radar channel estimation is performed, we obtain an unstructured measurement matrix $ \boldsymbol{ \mathcal{H}}  \in \mathbb{C}^{\Nsc \times \Nvirt} $ for $ \Nsc $ subcarriers and $ \Nvirt $ virtual spatial channels as defined in (15) in \cite{9589830}. In terms of column vectors, it is represented as
\begin{equation}
	\label{eqn_radar_measurement_column}
	\boldsymbol{ \mathcal{H}} = \big[ \bar{\textbf{h}}_{1,1}, \bar{\textbf{h}}_{1,2},  \mydots, \bar{\textbf{h}}_{\Nrx, \Ntx} \big], 
\end{equation}
where $ \bar{\textbf{h}}_{k,l} \in \mathbb{C}^{\Nsc}$ corresponds to the frequency response of the virtual spatial channel generated by $ l^{\text{th}} $ TX and $ k^{\text{th}} $ RX channels. Hence, each virtual channel vector is a summation of direct SI, static reflections, and targets' reflections which is formulated without channel indices as 
$$ \bar{\textbf{h}} = \bar{\textbf{h}}^{\mathrm{SI}} + \bar{\textbf{h}}^{\mathrm{TGT}}, $$
where $ \bar{\textbf{h}}^{\mathrm{SI}} $ contains the SI and background, and $ \bar{\textbf{h}}^{\mathrm{TGT}} $ is the target response. Assuming the targets are moving or quasi-static, the SI channel response is estimated with
\begin{equation}
	\label{eqn_si_mean}
	\bar{\textbf{h}}^{\mathrm{SI}}_{\mathrm{est}} = \frac{1}{N_{\mathrm{win}}} \sum_{i = -N_{\mathrm{win}}}^{-1} \bar{\textbf{h}}^{(i)},
\end{equation}
where $ N_{\mathrm{win}} \in  \mathbb{Z}_{+} $ is the depth of measurement window and $ \bar{\textbf{h}}^{(i)} $ is the $ i^{\text{th}}$ radar measurement where $ i = 0 $ corresponds to the latest measurement. Hence, the targets' channel is extracted from the latest radar measurement as
$
	\bar{\textbf{h}}^{\mathrm{TGT}}_{\mathrm{est}}  =  \bar{\textbf{h}}^{(0)} - \bar{\textbf{h}}^{\mathrm{SI}}_{\mathrm{est}}.
$
If the environment is dynamic, the SI estimation process should be repeated to track changes in the environment. Hence, we manually activate and deactivate the SI and background removal for our evaluations. The SI estimation frequency, automated activation, and distinguishing targets from the background are subjects of future work.

\section{Experimental Results}

In this section, we present experimental results to evaluate the JRC performance of the software-defined testbed at the 24 GHz band. For the evaluations, we conduct four different indoor experiments for (i) radar imaging with single and two targets (ii) radar path loss measurements, (iii) radar's angular power measurements, and (iv) communication path loss.

\begin{table}[!t]
	\centering
	\resizebox{0.9\columnwidth}{!}{
		\begin{threeparttable}
			\begin{tabularx}{\columnwidth}{c Y c}
				\hline \hline
				\textbf{Symbol} & \textbf{Parameter} & \textbf{Value} \\ \hline\hline
				$f_c$ & Carrier frequency & 24 GHz \\
				$B$ & Bandwidth & 125 MHz (200 MHz)\tnote{(1)} \\
				$ \Nsc $ & Number of subcarriers & 64 \\ 
				$ \Ntx $ & Number of transmit chains & 4 \\ 
				$ \Nrx $ & Number of receive chains & 2 \\ 
				\hline
				$ \rangeres $ & Range resolution & 1.20 m (0.75 m)\tnote{(1)} \\
				$ \angleRes $ & Angular resolution & 12.5$^{\circ}$ \\
				$ \rangemax $ & Maximum unambiguous  range & 76.8 m (48.0 m)\tnote{(1)} \\
				\hline
				$G_{\mathrm{tx}} $ & Gain of a TX chain  & 12.0 dB\tnote{(2)}  \\
				$P_{\mathrm{1dB}} $ & Maximum transmit power  & 21 dBm\tnote{(2)}  \\
				$G_{\mathrm{rx}} $ & Gain of a RX chain  & 6.2 dB\tnote{(2)}  \\
				$F_{\mathrm{noise}} $ & Cascaded noise figure  & 8.4 dB\tnote{(2)}  \\
				\hline
			\end{tabularx}
			\begin{tablenotes}
				\item[(1)] MIMO JRC supports up to 200 MHz bandwidth with real-time streaming.
			\end{tablenotes}
		\end{threeparttable}
	}
	\caption{Final system parameters of the JRC implementation.}
	\label{table_parameters_jrc_system}
\end{table}

\begin{figure}[!b]
	\centering
	\includegraphics[width=\columnwidth]{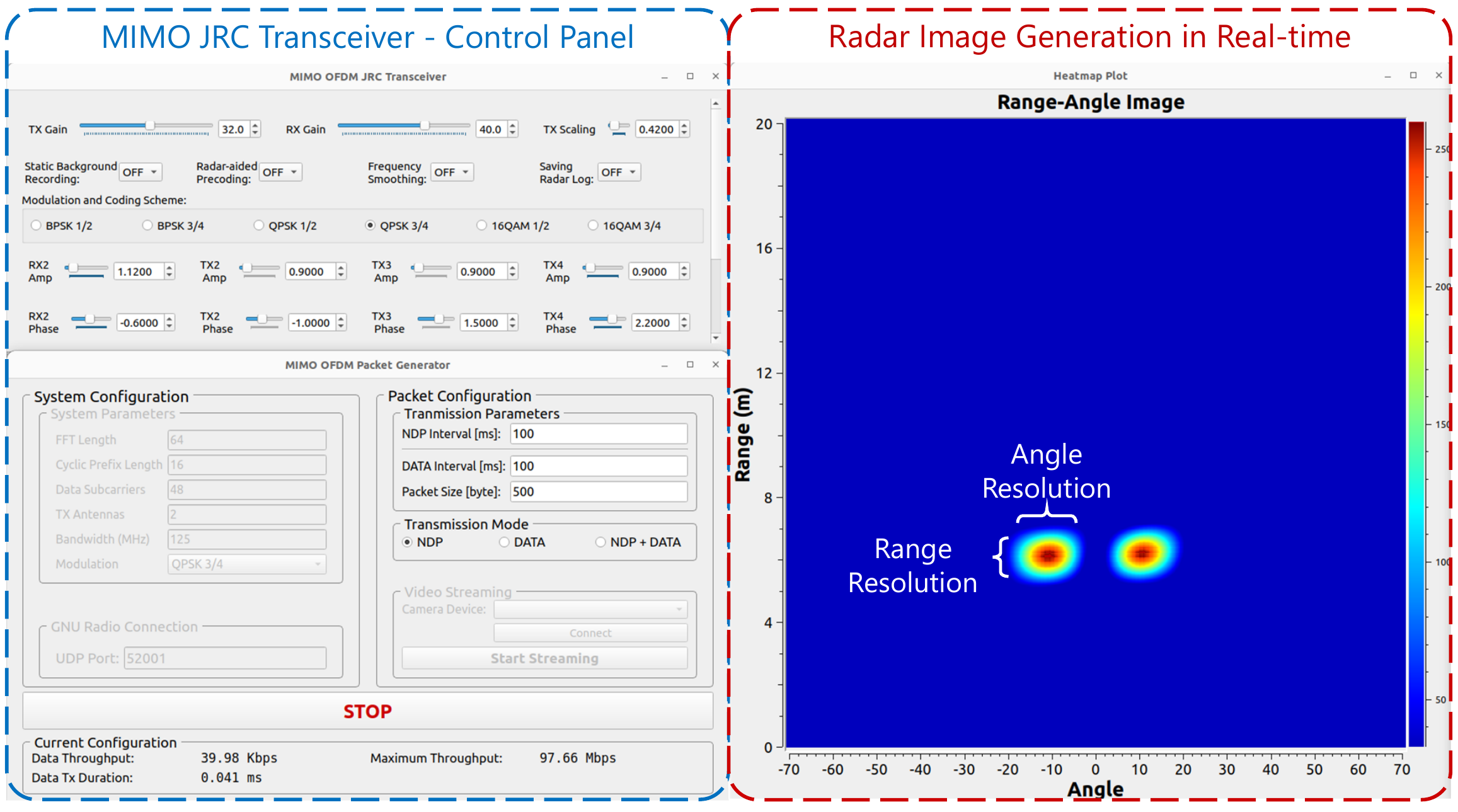} 
	\caption{The user interface of the MIMO JRC transceiver.}
	\label{fig_trx_gui_2_targets}
\end{figure}

We note that the MIMO JRC transceiver performs radar imaging operations whenever a transmission is initiated as it always operates in full-duplex mode. Once a packet is received from the upper layer via an internal UDP socket, the JRC transceiver flowgraph handles physical layer operations. Moreover, we implement packet generator software with a graphical user interface (GUI) that acts as an upper-layer application to communicate with the PHY flowgraph. The GUI of the MIMO transceiver is shown in Figure~\ref{fig_trx_gui_2_targets} in which the left window is a control panel for transceiver parameters and the right window displays range-angle images in real time. The summary of system parameters is given in Table~\ref{table_parameters_jrc_system}.

\begin{figure}[!b]
	\centering
	\includegraphics[width=\columnwidth]{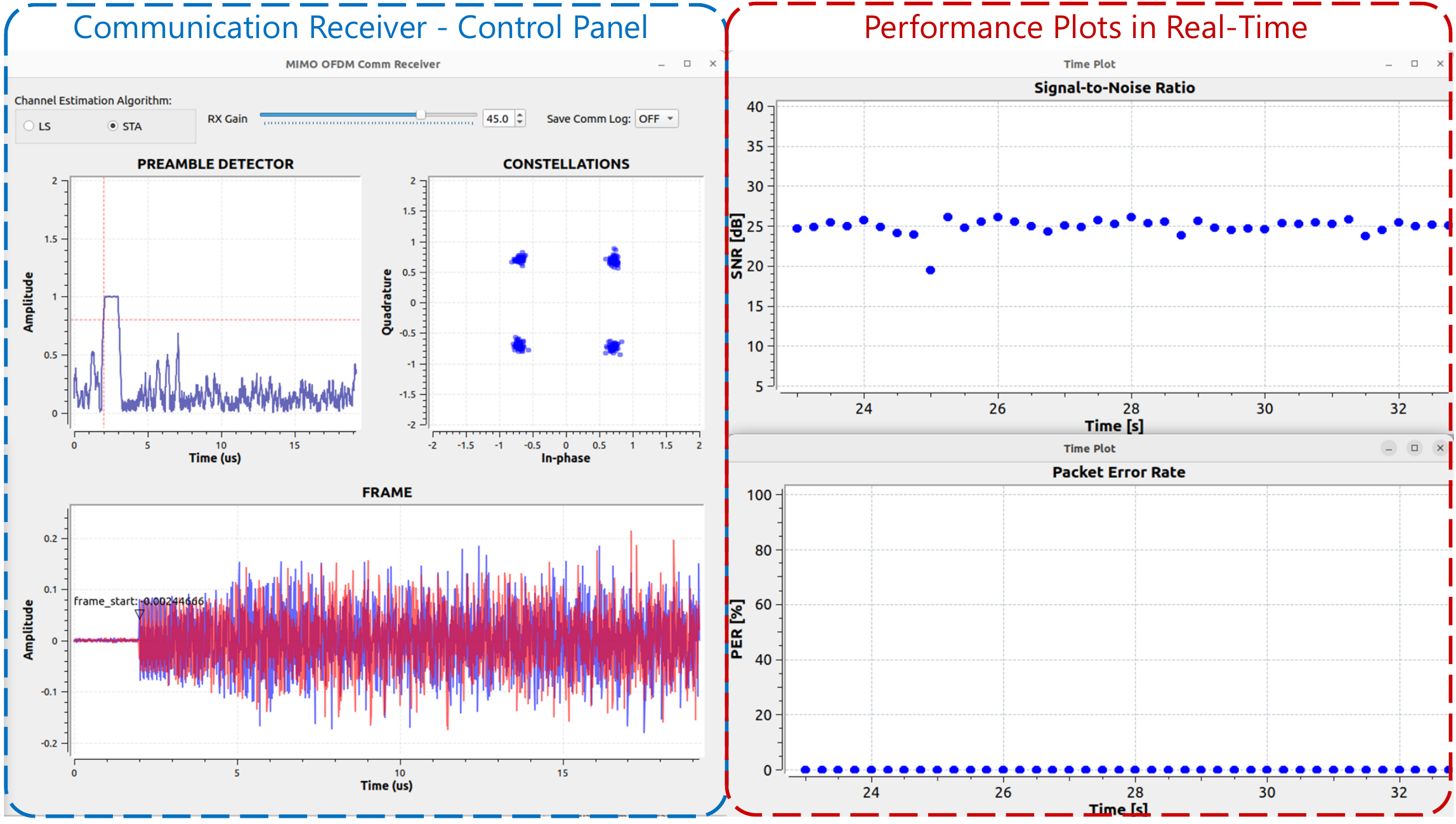} 
	\caption{The user interface of the communication receiver.}
	\label{fig_rx_gui}
\end{figure}

Also, a SISO transceiver is employed as a communication receiver that constantly executes the frame detection algorithm in real-time. Although the USRPs support 200 MHz bandwidth, the Host PC's processing rate can sustain only up to 125 MHz with the flowgraph of the communication receiver due to full-rate streaming. Hence, we present experimental results with 125 MHz bandwidth in this work. Nonetheless, the MIMO transceiver can perform radar processing with 200 MHz due to lower duty cycle reception with radar. Furthermore, the experiments are always conducted with real-time processing while instantaneous SNRs and packet error rates are shown in time plots. In Figure~\ref{fig_rx_gui}, the GUI of the communication receiver is shown with the output of the preamble detector, received signal, and constellations.

\subsection{MIMO Radar Imaging with Two Targets}

\begin{figure}[!t]
	\centering
	\includegraphics[width=\columnwidth]{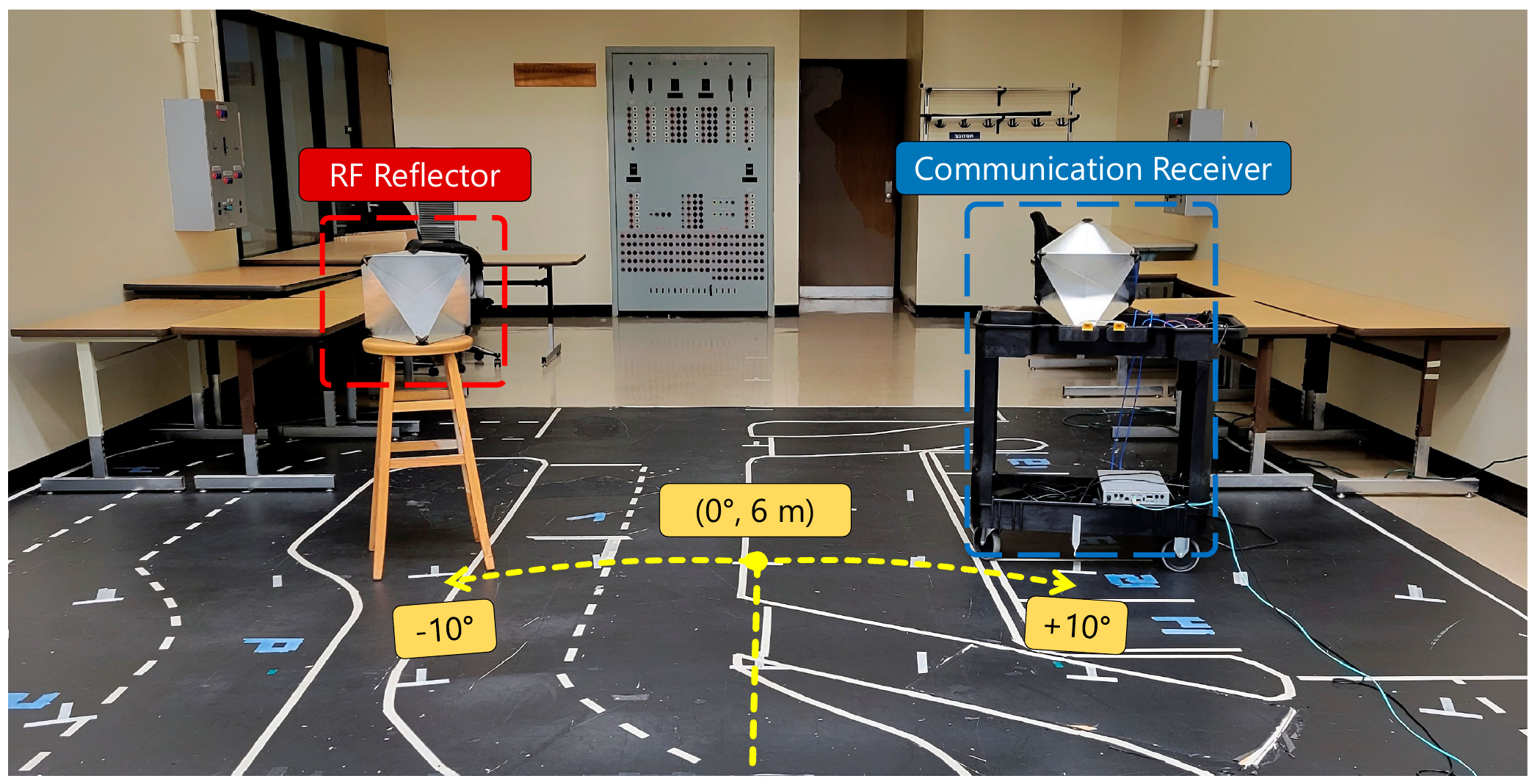} 
	\caption{Experiment scenario with a reflector at -10$ ^\circ $ and a mobile communication receiver at 10$ ^\circ $ with a horn antenna and a reflector.}
	\label{fig_radar_2targets}
\end{figure}

In this experiment scenario, we use two identical octahedral RF reflectors with an edge length of 34 cm. Without the reflectors in the scene, the proposed SI and background removal method is activated to cancel out static components from the radar image.  Then, we place two targets 6 m away from the transceiver at -10$  ^\circ  $ and $ 10 ^\circ  $ angles as shown in Figure~\ref{fig_radar_2targets}. For this scenario, Figure~\ref{fig_trx_gui_2_targets} shows the real-time display of the MIMO transceiver. As shown in the figure, the 3dB-widths of the target responses verify that the radar achieves expected range and angle resolutions of 1.2 m and 12.5$^\circ$, respectively.

\subsection{Path Loss Measurements with MIMO Radar} \label{section_radar_pl}

On the radar images, the peak value of a target response indicates the signal power whereas far-field values determine the noise floor. Hence, we compute the radar SNR by dividing the peak value by the noise floor level. Although the constant false alarm rate (CFAR) method is applied to detect multiple targets \cite{book_richards_radar_principles_2010}, we only extract the peak value for this scenario. 

\begin{figure}[!t]
	\centering
	\includegraphics[width=\columnwidth]{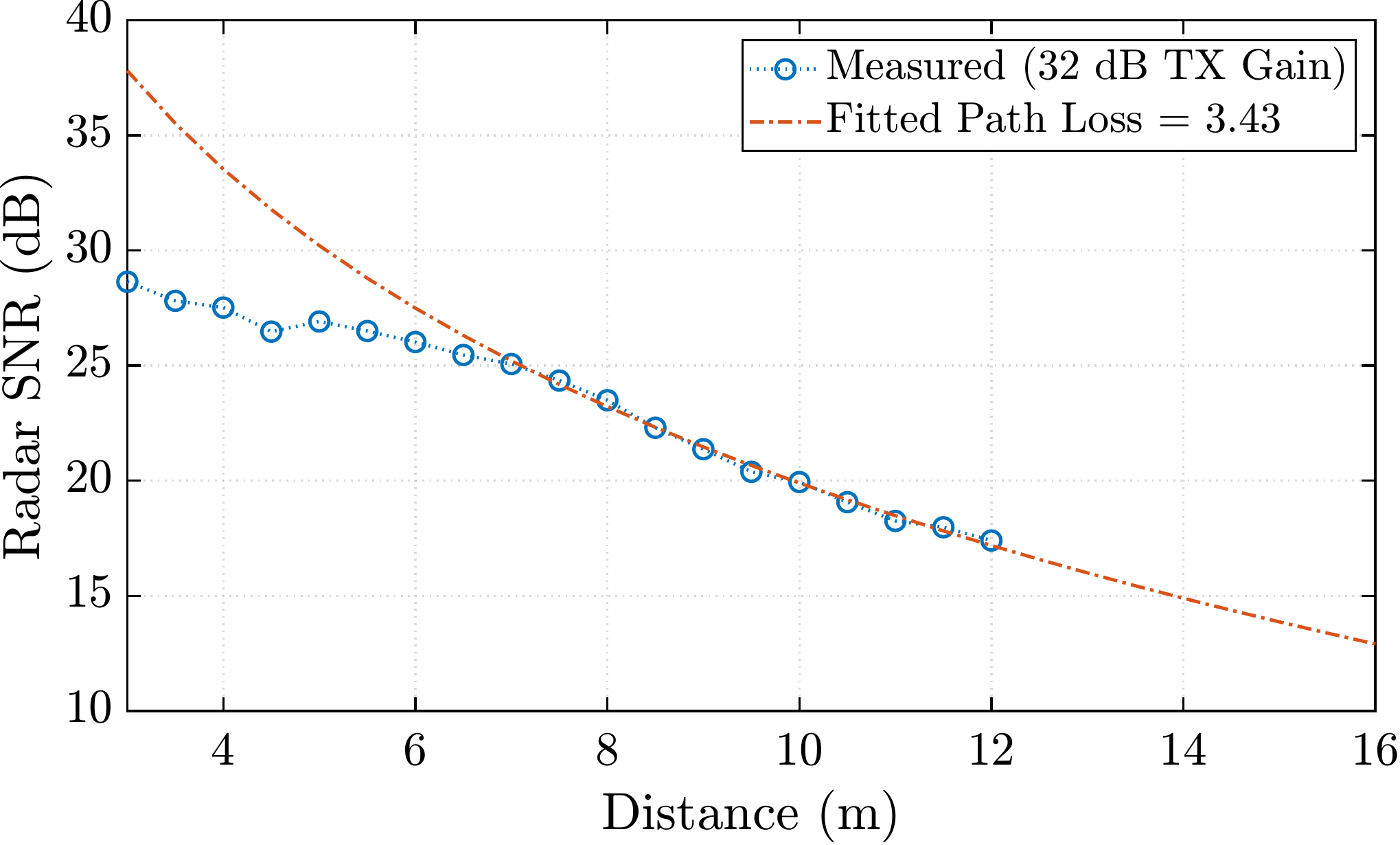} 
	\caption{Radar SNR measurements with the MIMO JRC platform when an RF reflector is moved from 3 m to 12 m away at 0$ ^\circ $ angle.}
	\label{fig_radar_path_loss}
\end{figure}

For this experiment, we set TX gain to 32 dB over 55 dB and move the reflector from 3 m to 12 m on 0$ ^\circ $ angle while the testbed is computing the radar SNR estimates. The change in the SNR estimate at different distances is shown in Figure~\ref{fig_radar_path_loss} where it achieves 17 dB SNR for a target 12 m away. To determine the path loss exponent, we define a curve-fitting problem with a path loss function that is formulated as $F_{\mathrm{PL}}(\beta_{\mathrm{PL}}, \alpha_{\mathrm{PL}}, d) = \beta_{\mathrm{PL}} - \alpha_{\mathrm{PL}} 10 \log_{10} \left(d/d_0\right),$ where $ \alpha_{\mathrm{PL}} $ is the path loss exponent, $ \beta_{\mathrm{PL}} $ is the function offset, $ d $ is the actual distance, and $ d_0 $ is a reference distance. For $ d_0 = 7 $ m, we solve a curve-fitting problem in least-squares sense using MATLAB's $ \mathsf{lsqcurvefit} $ function. 

As shown in Figure~\ref{fig_radar_path_loss}, we obtain a fitted path loss function with $  \alpha_{\mathrm{PL}} = 3.43 $, which is less than the theoretical two-way free-space exponent of 4. Hence, we conclude that the indoor mmWave radar channel introduces constructive reflections possibly from the floor and walls. Similar observations are also made in other experimental works  \cite{electronics9111867}. For the far-field, the radar SNR follows the fitted function. However, as the target comes closer, the radar's receive chains start to operate in the saturation region that results in gain compression.

\subsection{Angular Measurements with MIMO Radar}

\begin{figure}[!h]
	\centering
	\includegraphics[width=\columnwidth]{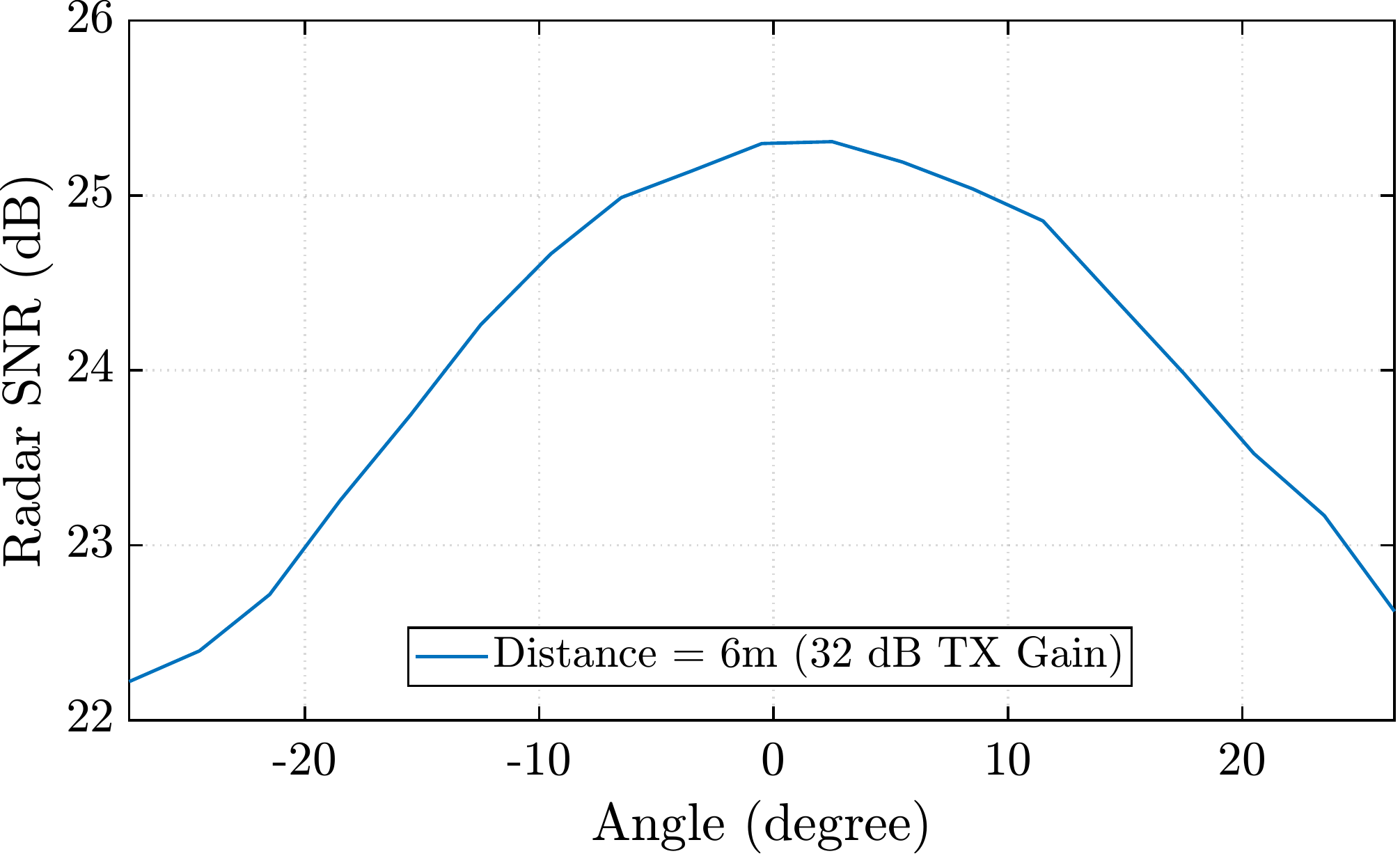} 
	\caption{Radar SNR measurements with the MIMO JRC platform when an RF reflector at 6 m is moved from -25$ ^\circ $ to +25$ ^\circ $ angles.}
	\label{fig_radar_snr_angle}
\end{figure}

For the angular measurements with the JRC platform, we logged radar SNRs while moving the reflector from -25$ ^\circ $ to +25$ ^\circ $ degrees at 6 m away. Figure~\ref{fig_radar_snr_angle} shows the measured SNR values at different angles. Although the theoretical composite beampattern of the MIMO radar is omnidirectional, the elemental gain factor of a patch antenna is never flat in practice. Nevertheless, we observe that the MIMO JRC platform achieves around 55$ ^\circ $ FoV with 3 dB loss as shown.

\subsection{Communication Path Loss Measurements}

In this section, we present the path loss performance of MISO communication where the MIMO testbed is the transmitter and the SISO testbed is the communication receiver. As shown in Figure~\ref{fig_radar_2targets}, the SISO platform is mounted on a rolling cart with a reflector for mobile experiments. While the MIMO JRC testbed transmits data-encoded waveform, its radar processor also tracks and records the position of the receiver during measurements.

\begin{figure}[!t]
	\centering
	\includegraphics[width=\columnwidth]{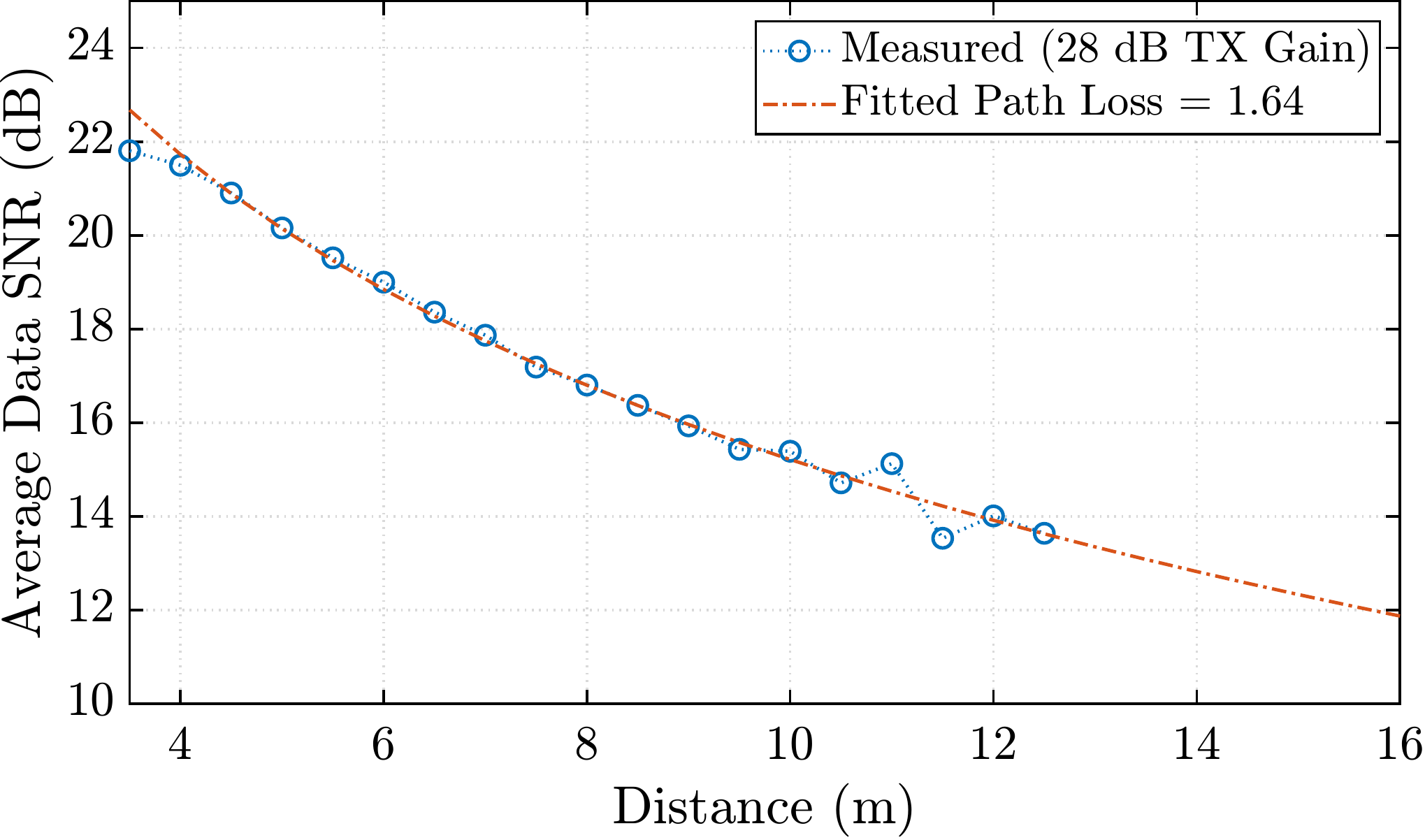} 
	\caption{Communication SNR measurements where the receiver moves from 3.5 m to 12.5 m away at 0$ ^\circ $ angle.}
	\label{fig_comm_path_loss}
\end{figure}

 During the measurements, precoded DATA packets with size of 500 bytes are transmitted every 100 ms and the communication SNR is estimated with precoded pilot symbols after equalization. To prevent gain compression at the receiver, we reduce the transmit gain to 28 dB. Then, we move the mobile receiver platform from 3.5 m to 12.5 m away. Figure~\ref{fig_comm_path_loss} shows estimated data SNR averaged over every 0.5 m distance with a fitted path loss function.  The estimated path loss exponent for communication is $  \alpha_{\mathrm{PL}} = 1.64  $ which is less than the theoretical exponent of 2 as also observed with radar's path loss measurements and in \cite{electronics9111867}.

\section{Conclusion}

In this work, we implemented a proof-of-concept $ 4 \times 2 $ MIMO OFDM JRC platform with fully digital architecture using SDRs that provide 200 MHz bandwidth. To operate in the 24 GHz band, we also designed custom mmWave front-ends using commercially available RF components.  For real-time range-angle imaging, we implement a low-complexity radar imaging algorithm for MIMO OFDM.  Moreover, we developed a self-interference and background removal approach for the MIMO-OFDM radar to improve target detection in the near-field. To demonstrate communication capabilities, we also built a SISO JRC platform mounted on a rolling cart. 

We conducted various experiments for different scenarios to characterize the capabilities of the MIMO JRC system. The radar imaging results showed that it can achieve high resolution via virtual array processing. Despite the latency and computational drawbacks of the SDR architecture, we show that the proposed MIMO OFDM-based JRC framework can be realized with USRPs for real-time testing. The SDR implementation of MIMO OFDM JRC makes it particularly easy to test, debug, and modify. To make the system accessible, we released our implementation under an Open Source license.

\bibliographystyle{IEEEtran}
\bibliography{references_short} 

\end{document}

%% file: mycommands.tex
\newcolumntype{Y}{>{\centering\arraybackslash}X}

\newcommand{\mydots}{\!...}
\DeclareMathSymbol{\shortminus}{\mathbin}{AMSa}{"39}

\newcommand{\rangeres}{\mathit{\Delta R}}

\newcommand{\angleRes}{\mathit{\Delta \theta}}

\newcommand{\rangemax}{\mathit{R}_{\mathrm{max}}}

\newcommand{\Ntx}{N_{\mathrm{tx}}}
\newcommand{\Nvirt}{N_{\mathrm{virt}}}

\newcommand{\Nrx}{N_{\mathrm{rx}}}

\newcommand{\Nsc}{N_{\mathrm{sc}}}

